\documentclass[%
 reprint,
 amsmath,amssymb,
 aps,
]{revtex4-1}
\usepackage{amsmath}
\usepackage{graphicx}
\usepackage{dcolumn}
\usepackage{bm}
\usepackage{hyperref}

\begin{document}

\preprint{APS/123-QED}

\title{Quasi-phase-matching of high harmonic generation using polarization beating in optical waveguides}

\author{Lewis Z. Liu}
 \email{L.Liu1@physics.ox.ac.uk}
\author{Kevin O'Keeffe}
\author{Simon M. Hooker}
\affiliation{%
 Clarendon Laboratory, University of Oxford Physics Department,
 Parks Road, Oxford OX1 3PU, United Kingdom
}%

\date{\today}

\begin{abstract}
A new scheme for quasi-phase matching high-harmonic generation is
proposed in which polarization beating within a hollow core
birefringent waveguide modulates the generation of harmonics. The
evolution of the polarization of a laser pulse propagating in a
birefringent waveguide is calculated and is shown to periodically
modulate the harmonic generation process. The optimum conditions for
achieving quasi-phase-matching using this scheme are explored and
the growth of the harmonic intensity as a function of experimental
parameters are investigated.

Please note that this is an arXiv version of the original APS paper.  Please cite original paper L. Z. Liu, K O'Keeffe, and S. M. Hooker, Phys. Phys. Rev. A 85, 053823 (2012).  APS link here: \url{http://pra.aps.org/abstract/PRA/v85/i5/e053823}
\begin{description}

\item[PACS numbers]
42.55.Vc 42.81.Gs 42.65.Ky

\end{description}
\end{abstract}

\maketitle


\section{Introduction}
High harmonic generation (HHG) is a nonlinear process in which odd
multiples of a fundamental driving field are produced when an
intense laser pulse is focused into a low density gas. HHG is an
attractive source of temporally and spatially coherent, tuneable
light with wavelengths in the XUV and soft X-ray range and has found
applications in areas such as time resolved measurements
\cite{Uiberacker, Schultze,Cavalieri}, ultrafast holography
\cite{RaHolography}, or diffractive imaging \cite{Sandberg}. A
semi-classical theory of HHG has been developed by Corkum
\cite{Corkum} and a quantum treatment has been given by Lewenstein
et al \cite{Lewenstien}.

Although HHG is an attractive source for a wide range of
experiments, its adoption in many applications is prevented by low
conversion efficiency, resulting in low signal strengths.  Typical
conversion efficiencies for HHG are approximately $10^{-6}$ at
photon energies around 100eV and $10^{-15}$ for photon energies
above 1keV. This low conversion efficiency is caused by a phase
mismatch between the driving laser field and the harmonics generated
at each point in the generating medium. This results in the
oscillation of the harmonic intensity with propagation distance,
with an amplitude of 100\%, preventing the continuous growth of the
harmonic field.  The period of this oscillation is $2L_c$, where
$L_c = \pi / \Delta k$ is the coherence length, and $\Delta k$
is the phase mismatch. Phase mismatch arises from neutral gas and plasma dispersion  and, if one is employed, the waveguide used to guide the driving radiation \cite{Murnane1999, Murnane1998}.

When generating harmonics in a hollow-core waveguide it is possible
to balance the dispersion due to the waveguide and free electrons with
that of the neutral atoms. In this case true phase-matching, $\Delta
k = 0$, can  be achieved resulting in quadratic growth of the
harmonic signal over extended regions. However, this approach is
limited to low levels of ionization, and consequently low laser
intensities, due to the dominance of the free electron dispersion at
higher ionization levels, placing a limit on the maximum harmonic
order which can be phase-matched using this technique
\cite{Murnane1999, Murnane1998}.  Other phase matching schemes include difference frequency mixing or angular tuning of crossing beams have been proposed \cite{ShkolnikovJOSAB, Birulin} 

An alternative approach to overcoming the phase mismatch is to
suppress HHG in out-of-phase regions, a technique known as
quasi-phase-matching (QPM). Various schemes have been developed for
quasi-phase matching HHG.  A series of gas jets,
appropriately-spaced, has been used to achieve QPM\cite{GasJet1,
GasJet2}, but with this technique the total number of zones is limited by the
Rayleigh range of the focused driving laser.  Most
quasi-phase matching schemes rely on guiding the driving laser pulse
in a hollow-core waveguide since this extends the region over which
harmonics can be generated. Experiments using a train of
counter-propagating pulses that represses or scrambles the harmonic
generation in the destructive zones have demonstrated
quasi-phase-matching over up to 5 zones of harmonics with photon energies in the range  \cite{Robinson2010, Dromey, Peatross, Lytle, Zhang}. Other QPM
schemes include using a corrugated waveguide \cite{ModWaveguide} or
multi-mode beating \cite{Zepf2007, RobinsonThesis} in hollow-core
waveguides.

In this paper we propose a new QPM technique -- polarization-beating QPM (PBQPM) -- which utilizes polarization beating in a
birefringent waveguide to modulate the generation of harmonics
\cite{Patent}. The key advantage of PBQPM is its simplicity, since it
avoids the need for additional laser pulses or
longitudinally-structured gas targets. Instead, the only requirement
is a birefringent waveguide with a suitable value of the
birefringence.

In Section \ref{Sec:PBQPM} the concept of PBQPM is outlined. Section
\ref{Sec:EnvFunc} develops the theory of PBQPM, and Section
\ref{Sec:Prop} explores the optimal parameters of this scheme.  In
Section \ref{Sec:Discussion}, birefrigent wavegudies and advantages
of and limits to PBQPM are discussed.

\section{Polarization Beating QPM \label{Sec:PBQPM}}

It is well known that the single-atom efficiency of HHG depends
sensitively on the polarization of the driving laser field
\cite{Budil1993, SolaPolarization, Antoine, Antoine2}, which arises from the fact that the ionized
electron must return to the parent ion in order to emit a harmonic
photon.

In PBQPM a birefringent waveguide is used to generate beating of the
polarization state of a driving linearly-polarized driving laser
pulse, thereby modulating the harmonic generation process. QPM will
occur if the period of polarization beating is suitably matched to
the coherence length of the harmonics.  In a birefringent waveguide,
the incident radiation can be resolved into two components polarized
along the birefringent axes. For linearly-polarized incident light
these components are initially in phase, but the difference in their
phase velocity will cause the two components to develop a phase
difference which increases linearly with propagation distance. As a consequence, the resultant polarization state will evolve from linearly polarized at an angle $\Theta$ to (for example) the fast axis, through (say) right-handed elliptical polarization to linearly polarized at an angle $ -\Theta$ to the fast axis, and thence through elliptical polarization
back to linearly-polarized radiation parallel to the incident light.

The polarization beat length $L_b$ is defined to be the distance for
the two polarization components to develop a phase difference of $\pi$, and
hence:
\begin{equation}L_b \equiv \frac{\pi}{\Delta \beta} =
\frac{\pi}{kB} = \frac{\lambda}{2B} \end{equation} where $\Delta
\beta$ is the difference in propagation constant for the two
polarization components, $\lambda$ is the vacuum wavelength, and $B =
\frac{\lambda \Delta \beta }{ 2 \pi} $ is the dimensionless
birefringence parameter

Thus, by matching the beat length to an appropriate multiple of
$L_c$, harmonic generation can be turned on and off along the length
of the waveguide in a controlled way -- enabling quasi-phase
matching.  Since the harmonics are generated most efficiently for linear polarization, the general condition for lowest-order for PBQPM is that the separation, $L_b$, of adjacent points of efficient harmonic generation is equal to an even number of coherence lengths, i.e. PBQPM requires, $L_b = n L_c$, where $n$ is an even integer  As is discussed in more detail below, since the coherence lengths for harmonics polarized parallel to the $\hat{x}$- and $\hat{y}$-axes are different, for a given $L_b$ it is only possible to quasi-phase-match one of these components.   As a consequence the output harmonics will be predominantly linearly polarised along the matched birefringent axis. Moreover, as discussed below, $\Theta$ and the harmonic order $q$ determine the width of the generation zone in relation to $L_b$ where $\Theta$ can be optimized.

\section{The envelope function for PBQPM \label{Sec:EnvFunc}}

If we write the electric field of the $q$th harmonic as
\begin{equation}E_q(z,t) = \xi(z,t) e^{i[k(q \omega)z - q \omega t]} \end{equation}
 then, within the slowly-varying envelope approximation:
\begin{equation} \frac{\partial \xi}{\partial z} = A \Lambda(z) e^{-i [k(q\omega) -
qk(\omega)]z}  \label{Eqn:Diff} \end{equation} where $\xi$ is the
electric field envelope, $k(\omega) = \beta(\omega)$ is the propagation constant for radiation of frequency $\omega$, $A$ is a normalization constant and
$\Lambda(z)$ is a relative source term.

For each polarization the wave vector mismatch for a specific harmonic, $q$, can be written as:
\begin{eqnarray}
\Delta k & = & k(q\omega) - qk(\omega)\\
             & = & \Delta k_{plasma} + \Delta k_{neutral} + \Delta
             k_{waveguide} \label{eqn:deltak}
\end{eqnarray}
and the envelope function can be solved as:
\begin{equation}\xi(z) = A \int_0^z dz' e^{-i\Delta k z'}
\Lambda(z') .\label{eqn:envfuncraw}
\end{equation}

\subsection{Effect of polarization state on HHG \label{Sec:Budil}}
The relative number of harmonic photons generated in the $q$th harmonic by a driving beam of ellipticity $\varepsilon$ may be written as:
\begin{equation} f(\varepsilon) \approx \left(\frac{1-\varepsilon^2}{1+\varepsilon^2} \right)^{\alpha} \label{eqn:f}\end{equation}
where $\varepsilon$ is defined as the ratio between the minor axis
to major axis.

Within the perturbative regime $\alpha = q-1$, as verified by Budil et al \cite{Budil1993} for harmonics $q = 11$ to $19$,  and by Dietrich et al  for harmonics up to  $q\approx 31$ \cite{DietrichPolarization}.  Schulze et al found that for higher-order harmonics the sensitivity of harmonic generation to the ellipticity of the driving radiation is lower than predicted by Eqn \eqref{eqn:f} with $\alpha = q-1$ \cite{SchulzePolarization}, although in this non-perturbative regime the efficiently of harmonic generation still decreases strongly with $\varepsilon$. Further measurements of the dependence of harmonic generation on ellipticity have been provided by Sola et al \cite{SolaPolarization}.   It is recognized that Eqn \eqref{eqn:f} is an approximation, but it will serve our purpose of demonstrating the operation of PBQPM. 

The offset angle and ellipticity of the harmonics
generated by elliptically-polarized radiation have been shown to
depend on the ellipticity and intensity of the driving radiation,
and on the harmonic order \cite{Soviet2, Antoine,  Antoine2, Strelkov, SchulzePolarization}.
Propagation effects can also play an important role. Since the
amplitude with which harmonics are generated decreases strongly with
increasing ellipticity, we are most interested in the ellipticity of
the harmonics generated for small $\varepsilon$. It has been shown
that for higher-order harmonics, and/or high driving intensities,
both the ellipticity and change in ellipse orientation of the
harmonics generated by radiation with $\varepsilon \approx 0$ are
close to zero \cite{Antoine}. We will therefore make the
simplification that the generated harmonics are linearly polarized along the major axis of the driving radiation, 
and that the harmonics polarized along the fast and slow axes of the
waveguide may be treated separately.

\subsection{Birefringence \& evolution of the ellipticity of polarization \label{Sec:Angle} }

In this section, the evolution of the driving field's ellipticity will be developed.  We assume azimuthal symetry of the modes.  Let $\hat{x}$ and $\hat{y}$ be the birefringent axes of the waveguide with the $\hat{x}$ axis being the slower axis such that $\beta_x < \beta_y$.  The driving field can be decomposed into the $\hat{x}$ and $\hat{y}$ components: \begin{equation} \vec{\mathfrak{E}}(r, z,t) =
\left(  \begin{array}{ll}
    \mathfrak{E}_x \\
   \mathfrak{E}_y \end{array} \right)  =
E(r)  \left(  \begin{array}{ll}
    e^{i([(\beta_y - \Delta \beta) z - \omega t]} \sin \Theta \\
    e^{i[ (\beta_y ) z - \omega t]} \cos \Theta \end{array} \right)
     \end{equation}
 where $E(r)$ describes the transverse electric field profile as a function of the distance $r$ from the propagation axis.

At a given point $z$ the electric field has an ellipticity given by:
\begin{equation} \varepsilon(z)  = \sqrt{ \frac{ \sin^2 \Theta \cos^2
(\phi_{min} - \Delta \beta z) + \cos^2 \Theta \cos^2(
\phi_{min})}{\sin^2 \Theta \cos^2 (\phi_{max} - \Delta \beta z) +
\cos^2 \Theta \cos^2 (\phi_{max})} }  \end{equation} where
\begin{equation} \tan^2 \Theta =- \frac{\sin [2 \phi]}{\sin[2(\phi +
\Delta \beta z)]} \label{eqn:maxmin} \end{equation} Here $\phi_{max}$ and $\phi_{min}$ are the values of $\phi =  \omega t$ which give respectively the maximum and minimum magnitude of the electric field.

\begin{figure}[tb]
\centering
\includegraphics[width=7cm]{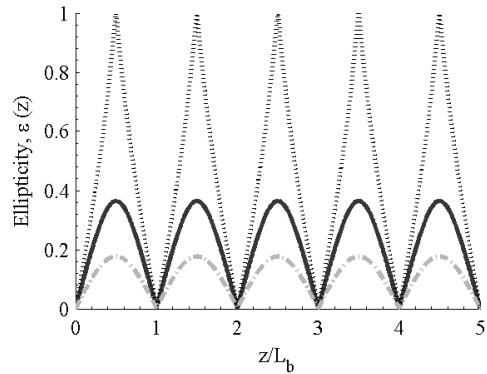}
\caption{Ellipticity beating for different incidence angles
$\Theta$. From top to bottom: Dotted light gray line shows $\Theta =
45^{\circ}$,  Solid gray line shows $\Theta = 30^{\circ}$ and
$\Theta = 70^{\circ}$; and dot-dashed gray line shows $\Theta =
20^{\circ}$ and $\Theta = 80^{\circ}$.}
\label{fig:ellipandaxislength}
\end{figure}

Although the ellipticity is periodic with period $L_b$, the electric
field is periodic with period $2L_b$.  The beating of the
ellipticity as a function of propagation distance for various
different angles of incidence is shown in Fig.
\ref{fig:ellipandaxislength}.

\subsection{Wave-vector mismatch in a birefringent waveguide}

In a birefringent waveguide the wave vector mis-match $\Delta k$ is
polarization-dependent.  However, the only polarization-dependent term in Eqn
\eqref{eqn:deltak}is $\Delta k_{waveguide}$, the wave vector mis-match arising from the waveguide dispersion. Hence we may write,
\begin{equation}\left\{ \begin{array}{ll}
    \Delta k_{x,waveguide} = \beta_x (q\omega) - q\beta_x (\omega)\\
    \Delta k_{y,waveguide} = \beta_y (q\omega) - q\beta_y (\omega) \end{array} \right. \end{equation}
From this, we may then write,
$$\Delta k_x = \Delta k_y + \delta k_{xy},$$
where,
\begin{align}
\delta k_{xy} &= \Delta k_{y,waveguide} - \Delta k_{x,waveguide}
\\  &= q \Delta
\beta(\omega) - \Delta \beta(q\omega) \\ &\approx q \Delta
\beta(\omega)
\end{align}
and the approximation follows if the birefringence is small at the
frequency of harmonic $q$.

\subsection{Constructing the PBQPM envelope function}

\begin{figure}[tb]
\centering
\includegraphics[width=7cm]{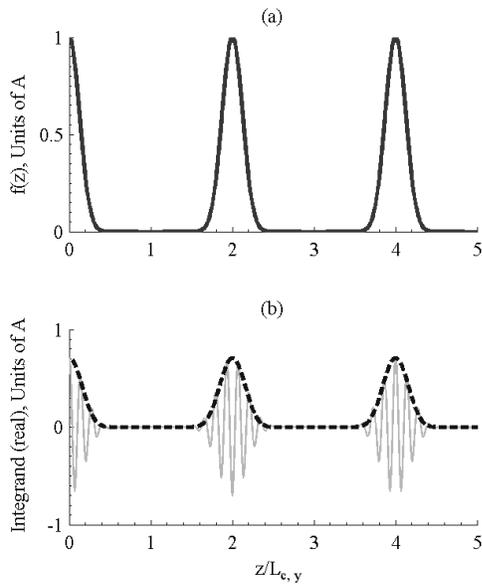}
\caption{Top graph (a): Variation of the harmonic generation efficiency, $f(z)=f[\varepsilon(z)]$, as function $z$
for $q=27$ and $\Theta = 45^{\circ}$. Bottom graph (b): Real part of
the of the $\hat{x}$ component of the integrand  (solid gray) and
$\hat{y}$ component  (dashed black line). Assuming $L_b =
2L_{c,y}$, $\Theta = 45^{\circ}$ and $q=27$}
\label{fig:HHGmodulation}
\end{figure}

From the discussion in Section \ref{Sec:Budil} we may treat each
polarization separately, and hence Eqn \eqref{eqn:envfuncraw} becomes:
\begin{equation}\left\{ \begin{array}{ll}
         \xi_x (z) = A  \int_0^z dz'e^{-i \Delta k_x z'} \Lambda_x(z') \\
         \xi_y (z) = A \int_0^z dz'e^{-i \Delta k_y z'} \Lambda_y(z') \end{array} \right. \end{equation}
where
\begin{equation}\left\{ \begin{array}{ll}
\Lambda_x(z') = \sqrt{f[\varepsilon (z')]} \sin\Theta \\
\Lambda_y(z') = \sqrt{f[\varepsilon (z')]} \cos\Theta
\end{array} \right. \end{equation} are the relative source terms.  In terms of the faster
polarization state (the $\hat{y}$ component),
\begin{equation}\left\{ \begin{array}{ll}
         \xi_x (z)= A \sin\Theta \int_0^z dz'e^{-i (\Delta k_y + q\Delta \beta) z'}  \sqrt{f[\varepsilon (z')]} \\
         \xi_y (z)= A \cos\Theta \int_0^z dz'e^{-i \Delta k_y z'} \sqrt{f[\varepsilon (z')]} \end{array} \right. \end{equation}

 Fig.\ \ref{fig:HHGmodulation} shows an example of $f[\varepsilon(z)]$ and the real part of the envelope function intergrand, assuming $\alpha = q-1$ and $L_b = 2L_{c,y}$.  Notice that integrating across the
$\hat{y}$ component would result in a monotonic increase of the
HHG amplitude whereas for the $\hat{x}$ component, no net increase in
harmonic amplitude would result owing to the rapid oscillations of
the integrand.  We conclude that the the polarization of the output
harmonic will be almost perfectly linear, in this case with an
electric field parallel to the y-axis since it is for that
polarization that the polarization beating has been matched to the
coherence length

\begin{figure}[tb]
\centering
\includegraphics[width=7cm]{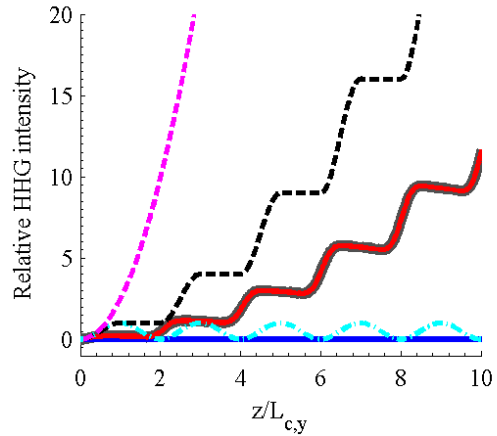}
\caption{Relative HHG intensity for different phase matching
conditions assuming that $L_b = 2 L_{c,y}$, $q=27$, $\Theta =
20^{\circ}$ for PBQPM. Dashed magenta line shows perfect phase
matching, dashed black line for perfect quasi-phase matching,
dot-dashed cyan line for no phase matching. The solid blue line
shows the relative amplitude squared for the $\hat{x}$ component whilst the
dashed yellow line shows the relative amplitude squared for the $\hat{y}$ component.  The thick gray line shows the total HHG
intensity for PBQPM.} \label{fig:HHGGainOneExample}
\end{figure}

Fig.\ \ref{fig:HHGGainOneExample} compares PBQPM for these
parameters against perfect phase matching and perfect QPM; we define the latter to correspond to square-wave modulation of the local harmonic generation with a period $2L_c$ and complete suppression of the out of phase zones.

\section{Properties of PBQPM\label{Sec:Prop}}
\subsection{Coherence length and beat length matching}

\begin{figure}[h]
\centering
\includegraphics[width=8cm]{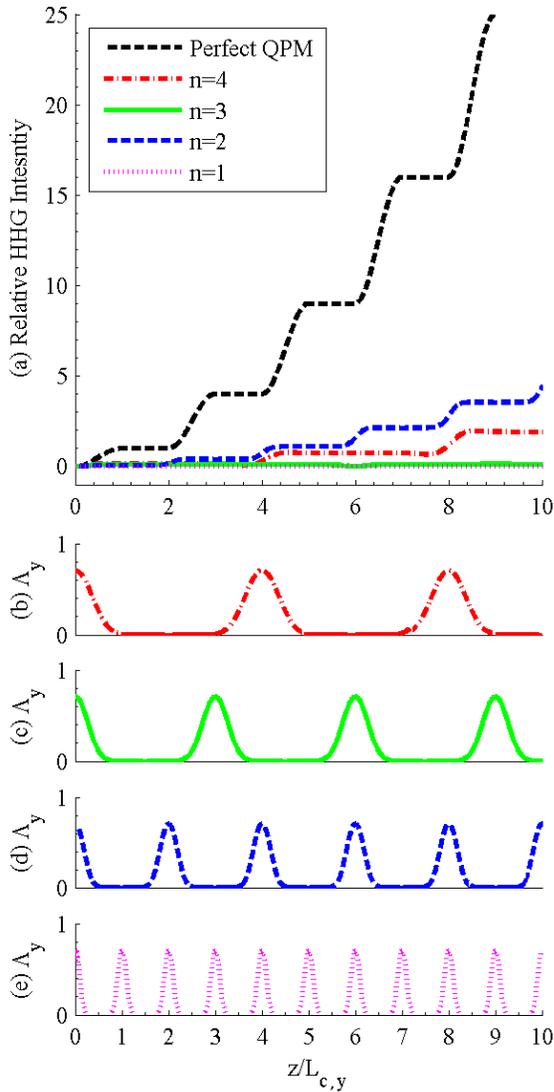}
\caption{PBQPM for $\Theta = 34^o$, $q = 27$, and $L_b = n L_{c,y}$ with
$ n= 1,2,3,4$. The top graph shows the relative harmonic intensity
as a function of propagation distance $z$ for perfect QPM (dashed
black line), $L_b=4L_{c,y}$ (dot-dashed red line), $L_b =3L_{c,y}$ (solid
green line), $L_b = 2L_{c,y}$ (dashed blue line), and $L_b = L_{c,y}$
(dotted magenta line). The bottom four graphs show the $p=y$
polarization source term $\Lambda_y$ as a function of propagation
distance $z$ for (b)$L_b=4L_{c,y}$,(c)$L_b=3L_{c,y}$, (d)$L_b=2L_{c,y}$, and
(e)$L_b=L_{c,y}$ } \label{fig:HHGGainDifferentLb}
\end{figure}

For a waveguide with fixed birefringence $B$ the condition for PBQPM can be realized by tuning the gas pressure and/or adjusting the driving laser intensity  until the period of polarization beating and the coherence length are appropriately matched. The general condition for QPM is that the distance over which harmonics are generated and that over which generation is suppressed are both equal to an odd number of coherence lengths, i.e. to $(2l+1)L_c$ and $(2m+1)L_c$ respectively, where $l$ and $m$ are integers. Thus the QPM period is in general $2(l+m+1)L_c = n L_c$ and hence the general condition for PBQPM is $L_b = n L_c$, where $n$ is even. Note that if the distances over which harmonics are generated and suppressed are equal, i.e. $l=m$, then PBQPM requires satisfaction of the more restrictive condition $L_b = 2(2l+1)L_c$. These considerations are illustrated by Fig. \ref{fig:HHGGainDifferentLb} which shows the calculated growth of the harmonic intensity for the cases $n=1,2,3$ and $4$. It may be seen that monotonic growth of the harmonics does not occur when $n$ is odd, as expected. Notice that PBQPM does occur for the case $n=4$, which corresponds to the lengths of the suppressed and unsuppressed regions of harmonic generation being unequal; this is possible for PBQPM since the efficiency of harmonic generation is very sensitive to the ellipticity, so that it is possible for the harmonic generation regions to be shorter than those in which generation is suppressed.

\subsection{Output intensity and coupling angle $\Theta$}
\begin{figure}[tb]
\centering
\includegraphics[width=8cm]{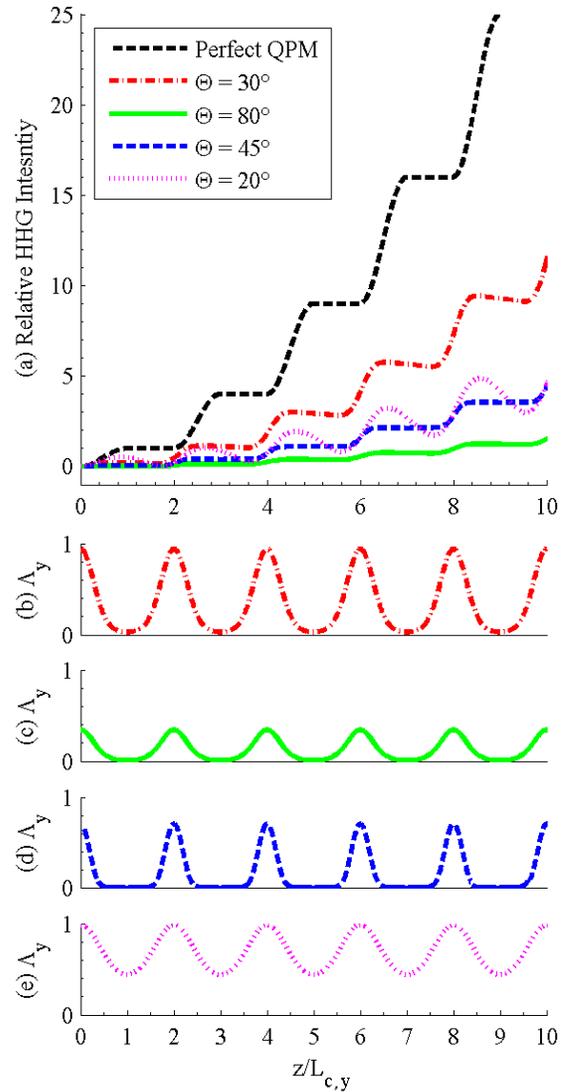}
\caption{PBQPM for $q=27$ and $L_b = 2L_{c,y}$ with $\Theta = 30^o,
80^o, 45^o, 20^o$.  The top graph shows the relative harmonic
intensity as a function of $z$ for perfect QPM (dashed black line)
and coupling angles of $\Theta = 30^{\circ}$ (dot dashed red line),
$\Theta = 80^{\circ}$ (solid green line), $\Theta = 45^{\circ}$
(dashed blue line), $\Theta = 20^{\circ}$ (dotted magenta line). The
bottom four groups show the $p=y$ polarization source term
$\Lambda_y$ as a function of $z$ for (b)$\Theta = 30^{\circ}$,
(c)$\Theta = 80^{\circ}$, (d)$\Theta = 45^{\circ}$, and (e) $\Theta
= 20^{\circ}$.} \label{fig:HHGGainDifferentTheta}
\end{figure}

\begin{figure}[tb]
\centering
\includegraphics[width=8cm]{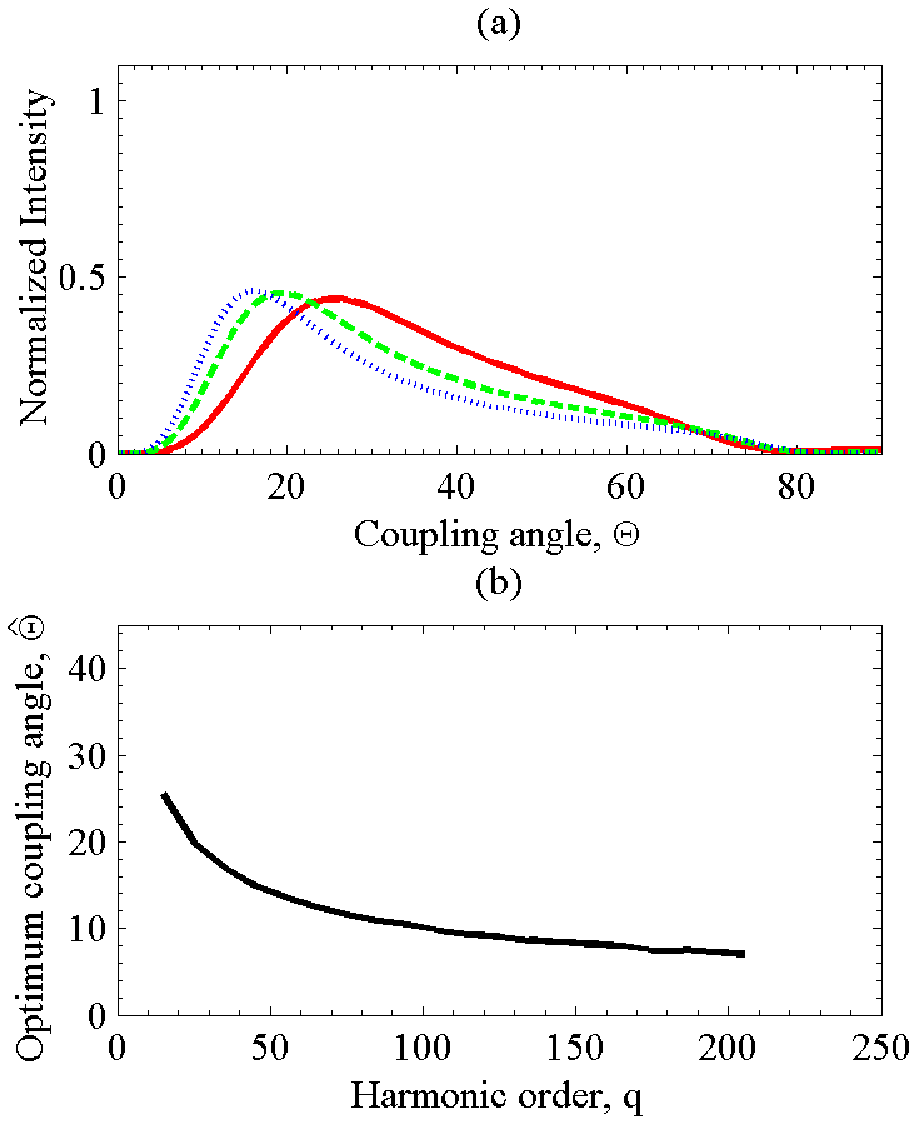}
\caption{Top graph (a): Relative HHG intensity as a function of
coupling angle $\Theta$, normalized to perfect QPM after one beat length where $L_b = 2L_{c,y}$ for
$q=15$ (solid red line), $q=27$ (dashed green line) , $q=39$ (dotted
blue line). Bottom graph (b): Optimal angle $\hat{\Theta}$ as a
function of harmonic order $q$ for $L_b = 2L_{c,y}$
\label{fig:GainVsTheta} \label{fig:ThetaHatvsQ} }
\end{figure}

It can be expected that the coupling angle $\Theta$ has an optimum:
if $\Theta$ is small then the $\varepsilon$ remains small for all $z$, and the
harmonic generation is not sufficiently suppressed in the out of
phase zones; if it is too large then harmonics are only generated
for a small fraction of the in-phase zones. Fig.\
\ref{fig:HHGGainDifferentTheta} shows the relative harmonic
intensity as a function of propagation distance for various coupling
angles $\Theta$.  Fig.\ \ref{fig:ThetaHatvsQ}a shows as a function
of $\Theta$, and for various harmonics $q$, the normalized harmonic
signal after one beat length. Also shown is the optimum
coupling angle as a function of harmonic order $q$.

The optimum coupling angle, $\hat{\Theta}$, will decrease with harmonic order since
higher-order harmonics are more sensitive to the ellipticity of the
driving radiation as discussed in Section \ref{Sec:Angle} , and
hence for these harmonics it is possible to couple more of the
polarization of the incident radiation along the polarization of the
harmonics to be generated by PBQPM. Fig. \ref{fig:ThetaHatvsQ}b
confirms this for the case of $L_b = 2L_{c,y}$.  

\section{Discussion \label{Sec:Discussion}}

Having introduced the concept of PBQPM and explored its operation in
theory, we now discuss how this quasi-phase-matching scheme might be
realized in practice. Several types of birefringent waveguides have
been developed to date: elliptical or rectangular waveguides
\cite{OkamotoBook}; waveguides made from birefringent materials
\cite{Marcuse}; waveguides with nonuniform refractive index
\cite{OkamotoBook}; photonic crystal fibers (PCFs) \cite{Chau2008,
Yang2011, NKTPhotonics}; waveguides constructed with tunable
piezoelectric material \cite{Tsia2008}; tunable stress-induced
birefringent waveguides \cite{OkamotoBook}. Many of these birefringent waveguides employ solid cores, and are therefore not suitable for PBQPM.  However, we note that (non-birefringent) gas-filled PCFs have been used to generate high harmonics \cite{HecklPCFHHG}.   Moreover, commercially available hollow core birefringent PCFs have been demonstrated with birefringence parameters as large as $B > 10^{-4}$ \cite{NKTPhotonics, RobertsBire}, corresponding to matching a coherence length of $\sim 1$mm, and solid-core PCFs with $B> 10^{-31}$ have been demonstrated \cite{MejiaBire}.  The birefringence of optical crystals can be  as large as $B \approx 0.4$ \cite{KatzBire}. Although the birefringence of hollow-core waveguides made from materials of this type would be only a fraction of that of the wall material, it seems likely that construction of hollow-core waveguides with sufficient birefringence to match coherence lengths in the range of a few hundred microns will be possible. The development and study of hollow-core birefringent waveguides of this type will form the basis of future work.

The principal advantage of PBQPM is the simplicity of the experimental set up: a single laser pulse correctly coupled into a birefringent waveguide. Other QPM techniques are likely to be limited by  several factors: for pulse-train QPM, the number of ultrafast pulses which can be generated; the precision with which corrugated waveguides or gas jet arrays can be manufactured; or, for multi-mode QPM, different damping rates for the two waveguide modes. PBQPM does not suffer from these difficulties. As for any HHG scheme limits will be set by absorption of the harmonics and variation of the ionization level -- and hence of the coherence length -- within the driving pulse; and as for other techniques employing a waveguide,  damping of the driving radiation may, in principle, limit the length over which harmonics will be generated. A potential limitation unique to PBQPM is  dispersion of the two polarization components: the slower polarization will lag the faster one until the two polarization states separate and polarization beating ceases. In principle this limitation could be overcome by reversing the birefringence after the two polarization components have slipped significantly; the limits imposed by all these mechanisms will be explored in future work. 

Finally we note that it is also possible to achieve PBQPM in
non-birefringent waveguides by exciting different-order waveguide
modes with nonparallel polarizations. In this arrangement the
different phase velocities of the modes will lead to polarization
beating in addition to the beating of the intensity which is
utilized in multi-mode QPM \cite{Zepf2007}. Indeed, driving PBQPM
with different-order waveguide modes increases the available
parameter space and could allow higher-order harmonics to be
quasi-phase-matched and greater efficiencies to be achieved.

\section{Conclusion}
We have proposed a new method for quasi-phase-matching HHG --
polarization-beating QPM (PBQPM) -- in which polarization beating in
a birefringent waveguide is employed to modulate the efficiency with
which harmonics are generated. Quasi-phase-matching may then be
achieved by suitably matching the period of polarization beating
with the coherence length.

We have developed a theoretical model of PBQPM and used this to
demonstrate that PBQPM  can indeed enhance the intensity with which
harmonics can be generated. This model was also used to explore the
optimum coupling angle for PBQPM and to show that this depends on
the harmonic order $q$.

 The authors would like to thank the EPRSC
for support through grant No. EP/GO67694/1. L. Liu would like thank
the James Buckee Scholarship of Merton College for its generosity and
David Lloyd for many fruitful discussions.

\section{APS Copyright Notice}
Copyright to the this unpublished and original article submitted by the author(s), the abstract forming part thereof, and any subsequent errata (collectively, the “Article”) is hereby transferred to the American Physical Society (APS) for the full term thereof throughout the world, subject to the Author Rights (as hereinafter defined) and to acceptance of the Article for publication in a journal of APS. This transfer of copyright includes all material to be published as part of the Article (in any medium), including but not limited to tables, figures, graphs, movies, other multimedia files, and all supplemental materials. APS shall have the right to register copyright to the Article in its name as claimant, whether separately or as part of the journal issue or other medium in which the Article is included.

\bibliography{PBQPM_BibFinal}

\end{document}